\documentclass[fleqn,preprint,5p]{elsarticle}

\usepackage{hyperref}
\usepackage{color}

\usepackage{amsmath, amssymb, amsfonts}
\usepackage{graphicx}

\journal{Physics Letters B}

\def\beq{\begin{equation}}
\def\eeq{\end{equation}}
\def\beqa{\begin{eqnarray}}
\def\eeqa{\end{eqnarray}}
\def\ben{\begin{enumerate}}
\def\een{\end{enumerate}}
\def\bit{\begin{itemize}}
\def\eit{\end{itemize}}

\def\lnb{\overline{\ln}}

\begin{document}

\begin{frontmatter}
\title{Analytical two-loop soft mass terms of sfermions in Extended GMSB models\tnoteref{t1}}

\tnotetext[t1]{This article is registered under preprint number: 1505.07443 [hep-ph]}

\author{Tomasz Jeli\'nski\corref{ca1}}
\ead{tomasz.jelinski@us.edu.pl}

\author{Janusz Gluza}

\cortext[ca1]{Corresponding author}

\address{
Institute of Physics, University of Silesia, Uniwersytecka 4, PL-40-007 Katowice, Poland
} 

\begin{abstract}
Analytical two-loop contributions to soft masses of sfermions are derived in the Extended GMSB (EGMSB) model with one superpotential coupling between matter and messenger superfields. 
Analytical results allows to study in detail the whole range of the ratio $F/M^2$ of the spurion $F$-term, $F$, to the messenger scale $M$. 
It is shown that if 
$F/M^2$ 
is of the order of $1$, 
then one- and two-loop contributions to soft masses are of the same magnitude and their interplay  leads to novel sfermion mass patterns.
\end{abstract}


\end{frontmatter}

\section{Introduction}

Supersymmetric models which involve GMSB mechanism \cite{Giudice:1998bp} are popular frameworks for analyzing phenomenology of BSM scenarios  mainly due to their flavour universality of soft SUSY breaking terms.  

In this letter, we consider in more details the issue of sfermion masses
in extended GMSB scenario. Let us start by reviewing situation within pure
GMSB scheme.

%
It is well-known that in such 
setups, 1-loop $A$-terms and 1-loop soft masses of sfermions $\widetilde{\phi}_i$ are zero at the messenger scale $M$. The first non-trivial contribution to soft  masses arise at 2-loop level and can be
written as \cite{Dimopoulos:1996gy,Martin:1996zb}:
\beq\label{m2g2}
(m^2_i)_{g,2}=2N\sum_r\left(\frac{g_r^2}{16\pi^2}\right)^2C^r_i(Mx)^2f_{g,2}(x)
\eeq
where $g_r$ are gauge couplings for $SU(3)$, $SU(2)$ and $U(1)$ groups, respectively.
$N$ is twice the sum of the Dynkin indices of the messenger fields, $C^r_i$ is the quadratic Casimir  related to $\widetilde{\phi}_i$, 
while $x=F/M^2$ is a dimensionless parameter\footnote{For simplicity, we assume that $M$ and $F$ are real.} which sets the relation between $F$-term of the spurion, $F$, and the messenger scale $M$. Finally, the function $f_{g,2}(x)$,
\beqa\label{f2g}
f_{g,2}(x)&=&\frac{1+x}{x^2}\left[\ln(1+x)-2\mathrm{Li}_2\frac{x}{1+x}+\frac{1}{2}\mathrm{Li}_2\frac{2x}{1+x}\right]\nonumber\\
&&+(x\to-x)
\eeqa
only 
mildly depends
on $x$ such that the overall scale of \eqref{m2g2} is set by $g^4_r(Mx)^2/(16\pi^2)^2$. This quantity is crucial for the phenomenology of GMSB models. To arrange masses of sfermions at the level of a few $\mathrm{TeV}$, such that sparticles are accessible at the LHC2 or at one of its successors, and to avoid fine-tuning of parameters, the scale $Mx$ has to be of the order of $10^5\,\mathrm{GeV}$. This allows us to consider the following relation between\footnote{The mass scale $Mx$ is very often written as $F/M$, see e.g. \cite{Giudice:1998bp} and sometimes denoted by $\xi$, see \cite{Jelinski:2011xe,Jelinski:2013kta,Jelinski:2014uba}.} $M$ and $x$ 
\beq\label{Mx}
Mx=c_{\xi}\times10^5\,\mathrm{GeV},
\eeq
where $c_{\xi}$ is a numerical constant of the order of $1$. 
Its precise value depends on the details of the model (e.g. 
$\tan\beta$,  $\mu$-term). 

However, the GMSB models are not 
flexible enough to fulfil all phenomenological restrictions and simultaneously retain \eqref{Mx}. For example, it is hard to accommodate for $125\,\mathrm{GeV}$ Higgs boson mass when sfermions are relatively light \cite{Draper:2011aa}. 
It turns out that the above mentioned issue can be alleviated within so-called Extended GMSB (EGMSB) models in which 
one allows for superpotential coupling between messenger and MSSM fields. Beside addressing enhancement of the Higgs boson mass, EGMSB models also provide reasonable framework for exploring non-minimal flavour violation scenarios within various GUT models \cite{Jelinski:2014uba,Albaid:2012qk,Shadmi:2011hs,Abdullah:2012tq,Calibbi:2013mka,Galon:2013jba,Evans:2015swa}. As a result of the presence of additional contributions to soft terms, EGMSB models exhibit rich and interesting phenomenology in which e.g. light stops can be naturally realized \cite{Jelinski:2011xe,Evans:2011bea,Kang:2012ra}. 

It is known \cite{Dine:1996xk, Giudice:1997ni} that in EGMSB models both 1- and 2-loop soft masses are generated by the messenger-MSSM superpotential couplings. The 1-loop masses are known analytically for any $0<x<1$ and one can show that they are negative and $\propto h^2x^2(xM)^2/16\pi^2$ for $x\ll1$. 
On the other hand only the leading term in $x=F/M^2$ of the 2-loop contribution to soft masses is known. It has been extracted with the help of   
the wave-function renormalization method \cite{Jelinski:2013kta,Giudice:1997ni, Chacko:2001km,Joaquim:2006uz,Joaquim:2006mn,Evans:2013kxa}, which captures only contributions of the order of $(Mx)^2$. One can show that for $x\ll1$ the 2-loop contributions to soft masses are positive and $\propto h^4(xM)^2/(16\pi^2)^2$. 
%
Naively, taking into account above considerations one expects that 1-loop contributions could start to dominate for $x\gtrsim h^2/16\pi^2\sim10^{-2}-10^{-3}$.
Though, such rough estimation 
does not take into account higher powers of $x$ in the expansion of 2-loop masses, but also
does not encounter numerical factors\footnote{These numerical factors also enter 1-loop anomalous dimensions of chiral superfields \cite{Giudice:1997ni,Evans:2013kxa,Martin:1993zk}.} which originate from 
contracting tensors related to the representations of chiral superfields  
\cite{Giudice:1997ni,Evans:2013kxa}.

Let us recall that in many GUT models the messenger scale is chosen rather large,\footnote{In the GMSB models gravitino mass is $m_{3/2}\sim F/M_{Pl}=xM^2/M_{P}$, where $M_{P}\sim10^{18}\,\mathrm{GeV}$ is the Planck scale. If \eqref{Mx} is satisfied then $m_{3/2}\sim 10^{-8}\,\mathrm{GeV}/x$. Hence in the models with $x\sim1$ the gravitino mass is of the order of $10\,\mathrm{eV}$. Such light gravitino 
may be of some importance for the cosmology, see e.g. \cite{Yanagida:2012ef}.} e.g. $M=10^{14}\,\mathrm{GeV}$, what  due to \eqref{Mx} forces $x$ to be of the order of $10^{-9}$. As mentioned above for such choice of $x$ 
%
the full form of the loop function \eqref{f2g} is not necessary - only the leading term is relevant. 
However, there are also interesting setups, so-called Low-Scale Gauge Mediation models with relatively light messengers $M\sim10^5-10^6\,\mathrm{GeV}$ which provide interesting  phenomenology, see e.g. \cite{Craig:2012xp}.  In this situation $x$ can be much larger, even of order 1. In this regime wave-function method is not adequate.  
%
%
To provide more accurate estimation of soft masses, 
one needs full analytical 2-loop expression.

Our aim is to calculate such full analytical formula for 2-loop soft masses of sfermions which is valid for all $0<x<1$, examine its relation to 1-loop soft masses and check for which values of $x$ full analytical expressions are important. 
To derive them, 
we 
adopt a general method presented in \cite{Martin:2003qz,Martin:2003it}, which is suitable for 2-loop SUSY calculations, and conventions therein.

\section{EGMSB model with one messenger-matter coupling}
We consider a EGMSB-type model in which a pair of messengers, $Y$, $\overline{Y}$ couple to MSSM fields $\phi_1$, $\phi_2$ via superpotential term $hY\phi_1\phi_2$ and SUSY is broken by vev of $F$-term of spurion superfield $X$. 
In other words, the relevant part of the superpotential is of the following form:
\beqa\label{Wh}
W_h=XY\overline{Y}+h\phi_1\phi_2Y.
\eeqa
The model has symmetry $\phi_1\leftrightarrow\phi_2$ and, moreover, one can assign global $U(1)$ charges  to $\phi_i$ and $Y$, $\overline{Y}$ such that those fields cannot mix with each other even after SUSY breaking.\footnote{For the discussion of the role of such charges in the GUT models see e.g. \cite{Jelinski:2013kta,Heckman:2009mn,Dolan:2011iu}.} An example of such assignment is the following: $q_{\phi_i}>0$, $q_{\phi_1}\neq q_{\phi_2}$, $q_{\phi_1}+q_{\phi_2}=-q_Y=q_{\overline{Y}}$ and  $q_X=0$. Due to the above-mentioned symmetries: (i) both fields $\phi_{1,2}$ obtain the same soft mass $m^2$, (ii) there is no mixing mass term $\phi_1\phi_2^*+\mathrm{h.c.}$ in the scalar potential, and (iii) there is no mass term of the form $\phi_{i}\phi_j+\mathrm{h.c.}$ Finally, (iv) one can get rid of any complex phase of $h=|h|e^{i\phi_h}$ just by a redefinition of messenger fields: $Y\to Ye^{-i\phi_h}$, $\overline{Y}\to \overline{Y}e^{i\phi_h}$. Hence, without loss of generality, we can assume that $h$ is real.   
%
An example of superpotential interaction \eqref{Wh} is a coupling between up Higgs $\phi_1=H_u$, up quark $\phi_2=\overline{U}$ and messenger $Y=Y_Q$ in the representation $(3,2)_{1/6}$ of $SU(3)\times SU(2)\times U(1)$.

When spurion field gets vev $\left<X\right>=M+\theta^2F$, all components of messenger multiplets obtain masses $M$, while scalar components get extra SUSY breaking masses $FY\overline{Y}+\mathrm{h.c}$.  After unitary rotation of the scalar components of messengers $(Y,\overline{Y})$ to the basis $(Y_+,Y_-)$ in which their mass terms are diagonal, the scalar potential is of the form:
\beqa
V&=&M^2_+|Y_+|^2+M^2_-|Y_-|^2\nonumber\\
&&+\left[\frac{1}{\sqrt{2}}hM\phi_1\phi_2(Y_++Y_-)+\mathrm{h.c.}\right]\nonumber\\
&&+\frac{1}{2}h^2(|\phi_1|^2+|\phi_2|^2)|Y_+-Y_-|^2,
\eeqa
where $M_\pm ^2=M^2(1\pm x)$ and $x=F/M^2$. Let us remark that $x$ has an upper bound, related to the upper bound on the mass of $Y_-$, i.e. on $M_-=M\sqrt{1-x}$. Because $Y_-$ can have the same quantum charges as quark superfields hence LHC bound on squarks masses $m_{\mathrm{LHC}}\approx2\,\mathrm{TeV}$ \cite{Aad:2014wea,Khachatryan:2015vra} apply to $Y_-$ as well. 
However, for the simplicity, in this letter we assume a little more conservative bound:
\beq\label{Mminus}
M_-\gtrsim10^4\,\mathrm{GeV},
\eeq
such that $Y_-$ is significantly heavier than all sfermions. Otherwise $Y_-$ may, in some cases, 
alter stability of the scalar potential, what in turn can be dangerous for EWSB. 
Using definition of $M_-$, the condition \eqref{Mminus} can be rewritten in the following form:
\beqa\label{xbound}
x\lesssim1-\left(\frac{10^4\,\mathrm{GeV}}{M}\right)^2.
\eeqa
It ensures that below the scale of $10\,\mathrm{TeV}$ one gets only MSSM fields.   
For example, when $M=10^5\,\mathrm{GeV}$, $x$ has to be smaller than about $0.99$, but for $M=10^{14}\,\mathrm{GeV}$, that limit is practically equal to $1$ and may indicate a kind of severe fine-tuning of parameters: $x\lesssim1-10^{-20}$. 
%
Additionally, taking into account \eqref{Mx} the bound \eqref{xbound} can be rewritten as:
\beqa\label{xbound2}
x\lesssim1-\frac{1}{100c_{\xi}^2}.
\eeqa

Let us summarize our discussion. For $x\ll1$, the 2-loop soft masses can be obtained with the help of the wave-function renormalization method. We also know that the regime $x\lesssim1$ is interesting for the phenomenology of Low-Scale Gauge Mediation models. But for $x\sim1$ the wave-function method is not adequate. Hence, to fill that gap, we shall discuss full analytical result for 2-loop masses generated by \eqref{Wh}.

\section{1-loop soft-terms}

For the completeness, we shall present 1-loop $A$-terms and 1-loop soft masses in the discussed model. We omit soft masses of gauginos because at the 1-loop level  they are of the same form as in the standard GMSB models \cite{Giudice:1998bp}. The 1-loop soft mass of $\phi_i$ is
\beq\label{m2h1}
m^2_{h,1}=\frac{h^2}{16\pi^2}\left(-\frac{1}{6}\right)(Mx)^2x^2f_{h,1}(x),
\eeq
where the function $f_{h,1}(x)$ is defined as follows:
\beqa\label{fh1}
f_{h,1}(x)&=&\frac{3}{x^4}\left[(x-2)\ln(1-x)-(x+2)\ln(1+x)\right]\nonumber\\
&=&1+\frac{4}{5}x^2+\frac{9}{14}x^4+\ldots.\quad (\textrm{for $x\ll1$})
\eeqa
On the other hand the $A$-term is:
\beqa\label{A1}
A=-\frac{h^2}{16\pi^2}2Mxf_{h,A}(x),
\eeqa
where
\beqa
f_{h,A}(x)&=&\frac{1}{2x}\ln\frac{1+x}{1-x}\nonumber\\
&=&1+\frac{1}{3}x^2+\frac{1}{5}x^4\ldots.\quad (\textrm{for $x\ll1$})
\eeqa
It is clear that both $m^2_{h,1}$ and $A$ diverge when $x\to1$. The reason for that is the massless scalar $Y_-$ which is present in the 1-loop diagrams \cite{Malinsky:2012tp} and leads to the divergence of the following contribution $(M^2h^2/16\pi^2)\ln(M^2_-/M^2)$ to the self-energy $\Pi^{(1)}(0)$ at zero momentum $s=p^2=0$. 
To ensure that the perturbative expansion in $h^2/16\pi^2$ does not break down, the following conditions has to be satisfied
\beq\label{pert}
|1-x|\ll\exp\left(\frac{16\pi^2}{h^2}\right). 
\eeq
To find 1-loop pole mass $m^2_*$ of $\phi_i$ at $x=1$ one could solve the full propagator equation $m^2_*-\Pi^{(1)}(m^2_*)=0$ numerically.  

The discussed behaviour of 
\eqref{m2h1} and \eqref{A1} 
is not a big problem
because phenomenology restricts $x$ to the regime \eqref{xbound2} which is safe from divergence.  
%



\section{2-loop self-energy}
To compute 2-loop contributions to soft masses generated by \eqref{Wh}, we apply method developed in \cite{Martin:2003qz,Martin:2003it}. For $\phi_i$, we calculate 1-loop self energy $\Pi^{(1)}(s)$ at $s=p^2$ and 2-loop self energy $\Pi^{(2)}(0)$ at $s=0$ and use them to find 
the pole mass of $\phi_i$. One can show that 1- and 2-loop pole masses are of the following form:\footnote{The tree-level mass of $\phi_i$ is zero.}
\beqa\label{m21h}
m^2_{h,1}&=&\frac{1}{16\pi^2}\Pi^{(1)}(0),
\eeqa
\beqa\label{m22h}
m^2_{h,2}&=&\frac{1}{(16\pi^2)^2}\left[\Pi^{(1)}{'(0)}\Pi^{(1)}(0)
+\Pi^{(2)}(0)\right].
\eeqa
%
%
We can compute contributions to $\Pi^{(2)}(0)$ from Feynman diagrams with topologies $(W,S,\ldots)$ shown on Fig.~2  in \cite{Martin:2003it}:
\beqa\label{Pi2}
\Pi^{(2)}(0)&=&\Pi_{W_{SSSS}}+\Pi_{X_{SSS}}+\Pi_{Y_{SSSS}}+\Pi_{Z_{SSSS}}\nonumber\\
&&+\Pi_{S_{SSS}}+\Pi_{U_{SSSS}}+\Pi_{V_{SSSSS}}+\Pi_{W_{SFF}}\nonumber\\
&&+\Pi_{V_{SSSFF}}+\Pi_{V_{FFFFS}}
\eeqa
where $\Pi_{W_{SSSS}}$, $\Pi_{S_{SSS}}$,\ldots correspond to topologies $(W,S,\ldots)$ respectively. 
Structure of these functions is displayed in the Appendix. We have regulated 
IR divergencies of the components of \eqref{Pi2} by adding a mass term $m^2_{\mathrm{IR}}$ to all massless sfermions.\footnote{Higgs bosons are assumed to be massless for the simplicity. An example of messenger-matter coupling for which Higgs mass term is not relevant is $Y_{H_u}Q\overline{D}$ or $Y_{H_d}L\overline{E}$, where $Y_{H_{u,d}}$ are Higgs-type messengers.} The final result is finite in the limit $m^2_{\textrm{IR}}\to0$.  

We computed all the ingredients in 
\eqref{Pi2} 
and expressed them via 1-loop integrals\footnote{$\widehat{B}(x,y)$ is the standard $B(x,y)$ 1-loop function calculated at $s=0$.}
\beq
A(x)=x(\lnb x-1),\quad \widehat{B}(x,y)=-\int\limits_0^1 dt\lnb[tx+(1-t)y],
\eeq
the 2-loop integrals
\beq
I(x,y,z)=C^2\int d^dk \int d^d q\frac{1}{[k^2+x][q^2+y][(k+q)^2+z]},
\eeq
and derivatives of $I(x,y,z)$. $C=16\pi^2\mu^{2\epsilon}/(2\pi)^d$, $\mu$ is the regularization scale in $\overline{\mathrm{DR}}'$ scheme and $\epsilon=4-2\epsilon$. Finally, $\lnb x=\ln(x/Q^2)$, where $Q^2=4\pi e^{-\gamma}\mu^2$ is the renormalization scale.  
%
Due to supersymmetry, all $1/\epsilon^2$ and $1/\epsilon$ terms cancel from $m^2_{h,2}$ and so do all diagrams involving counter-terms.   

\begin{figure}[h!]
\begin{center}
\includegraphics[scale=0.75]{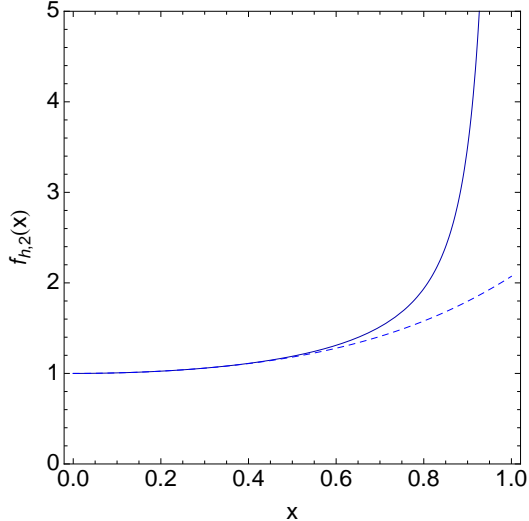}
\caption{The complete 2-loop function $f_{h,2}(x)$ defined in \eqref{m2h2} and \eqref{fh2} (solid line) and its approximation, \eqref{fh2x0}, valid for $x\ll1$ (dashed line). Note that $f_{h,2}(x)$ diverges for $x\to1$.}\label{fh2plot}
\end{center}
\end{figure}

Using explicit expressions for $I(x,y,z)$ we find that the 2-loop contribution to soft mass of $\phi_i$ generated by \eqref{Wh} at the renormalization scale $Q=M$ is of the following form:
\beq\label{m2h2}
m_{h,2}^2(M)=\left(\frac{h^2}{16\pi^2}\right)^23(Mx)^2f_{h,2}(x),
\eeq
where $f_{h,2}(x)$ is the complete analytical 2-loop function defined in the Appendix. 
The expansion of $f_{h,2}(x)$ for $x\ll1$ is:
\beq\label{fh2x0}
f_{h,2}(x)=1+\frac{1}{18}(\pi^2+1)x^2+\frac{1}{540}(30\pi^2-43)x^4+\ldots
\eeq
One can check that: (i) $f_{h,2}(x)$ is positive for $0\leq x<1$, (ii) $f_{h,2}(x)$ is divergent at $x=1$, analogously to \eqref{fh1}, and (iii) the numerical factor of $3$ in \eqref{m2h2} agrees with the corresponding numerical factor obtained using wave-function renormalization method \cite{Giudice:1997ni}. The plot of the function $f_{h,2}(x)$ is shown on the Fig.~\ref{fh2plot}, where $f_{h,2}(x)$ is compared with its approximation \eqref{fh2x0}. For $x\lesssim0.5$ the difference between   \eqref{fh2x0} and the full result is smaller than $1\%$.  

\section{
Analytical results: Consistency checks and phenomenological consequences
}

To verify our results we have made two basic consistency tests. Namely, we have checked that: (i) in the limit $x\to0$ (SUSY limit) the obtained 2-loop contribution  to soft-masses \eqref{m2h2} is zero, and (ii) the sum of 1- and 2-loop pole masses $m^2_{h}(Q)=m^2_{h,1}(Q)+m^2_{h,2}(Q)$ does not depend on the renormalization scale $Q$, i.e.
\beq
Q\partial_Qm_{h}^2(Q)=0 
\eeq
up to terms of order $O(h^5)$. To this end we have used the following 1-loop $\beta$ functions \cite{Martin:1993zk} for the running parameters: the messenger mass $M(Q)$, $F$-term $F(Q)$ and superpotential coupling $h(Q)$:
\beqa
\beta_{M}^{(1)}&=&h^2M,\\
\beta_{F}^{(1)}&=&-h^2F,\\
\beta_{h}^{(1)}&=&3h^3.
\eeqa
Let us remark that the term $\Pi^{(1)}(0)\Pi^{(1)}{'(0)}$ in \eqref{m22h} is crucial for the cancellation of $Q\partial_Qm^2_{2,h}$.

In addition, analytical formulas for integrals in \eqref{Pi2} have been cross-checked using numerical libraries in a wide range of $(\textrm{mass})^2$ parameters: from $10^{-13}\,\mathrm{GeV}^2$ to $10^{20}\,\mathrm{GeV}^2$. 
%
%
There are a few numerical libraries which can deal with 2-loop self energies:  \textsc{BOKASUN} \cite{Caffo:2008aw},  \textsc{CSectors} \cite{Bogner:2007cr,Gluza:2010rn}, \textsc{FIESTA} \cite{Smirnov:2013eza}, \textsc{SecDec} \cite{Borowka:2015mxa} and \textsc{TSIL} \cite{Martin:2005qm}. For our purpose the most appropriate turned out to be the \textsc{SecDec-3.0.6} with Suave routine from the \textsc{CUBA} library \cite{Hahn:2004fe}. 

\begin{figure}[h!]
\begin{center}
\includegraphics[scale=0.75]{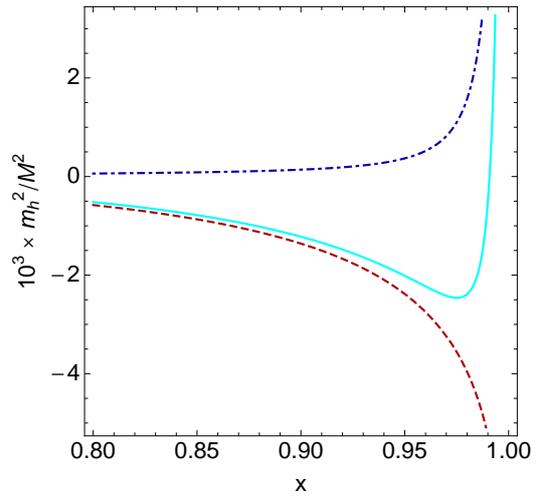}
\caption{Comparison of 2-loop (blue, dot-dashed) vs. 1-loop (red, dashed) contribution to soft mass for $x\sim1$. The superpotential coupling $h$ is fixed to $0.8$. The middle curve (solid) shows the sum of 1- and 2-loop contributions, $m^2_{h}=m^2_{h,1}+m^2_{h,2}$, generated by the messenger-matter coupling \eqref{Wh}.}\label{m2x1new}
\end{center}
\end{figure}

We have checked that due to interplay between 1- and 2-loop contributions to soft masses the apparent dominance of 1-loop contributions is not true for $x\sim1$ and it turns out that for $x\sim0.9$ the total contribution is bounded from below, see Fig. \ref{m2x1new}, instead of having a running direction caused by the 1-loop contribution.



One can check that the position of the minimum of $m^2_{h,2}$ on Fig.~\ref{m2x1new} only mildly depends on the value of $h$, e.g. increasing $h$ from $0.8$ to about $1.2$ changes its position from 0.975 to 0.96.  

\begin{figure}[h!]
\begin{center}
\includegraphics[scale=0.75]{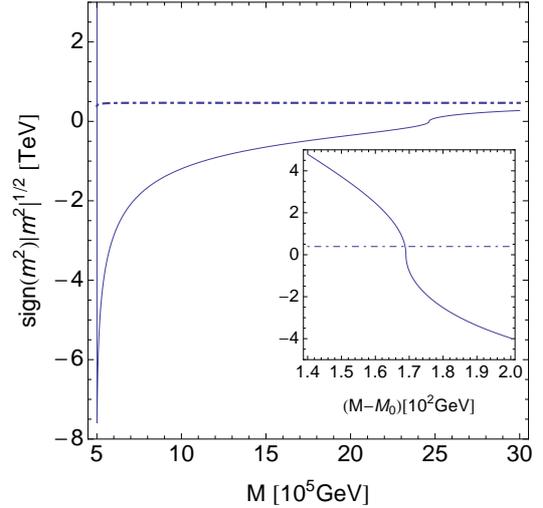}
\caption{Signed mass $\mathrm{sign}(m^2)\sqrt{|m^2|}$ of right-handed stau $\widetilde{\tau}_R$ as a function of the messenger scale $M$ for $h=0.1$ and fixed $Mx=5\times10^5\,\mathrm{GeV}$. The subplot displays the behaviour of signed mass for $M$ close to $M_0=5\times10^5\,\mathrm{GeV}$. The dot-dashed line represents mass of $\widetilde{\tau}_R$ in pure GMSB scenario (i.e. $h=0$).}\label{mseR3h01}
\end{center}
\end{figure}

\begin{figure}[h!]
\begin{center}
\includegraphics[scale=0.75]{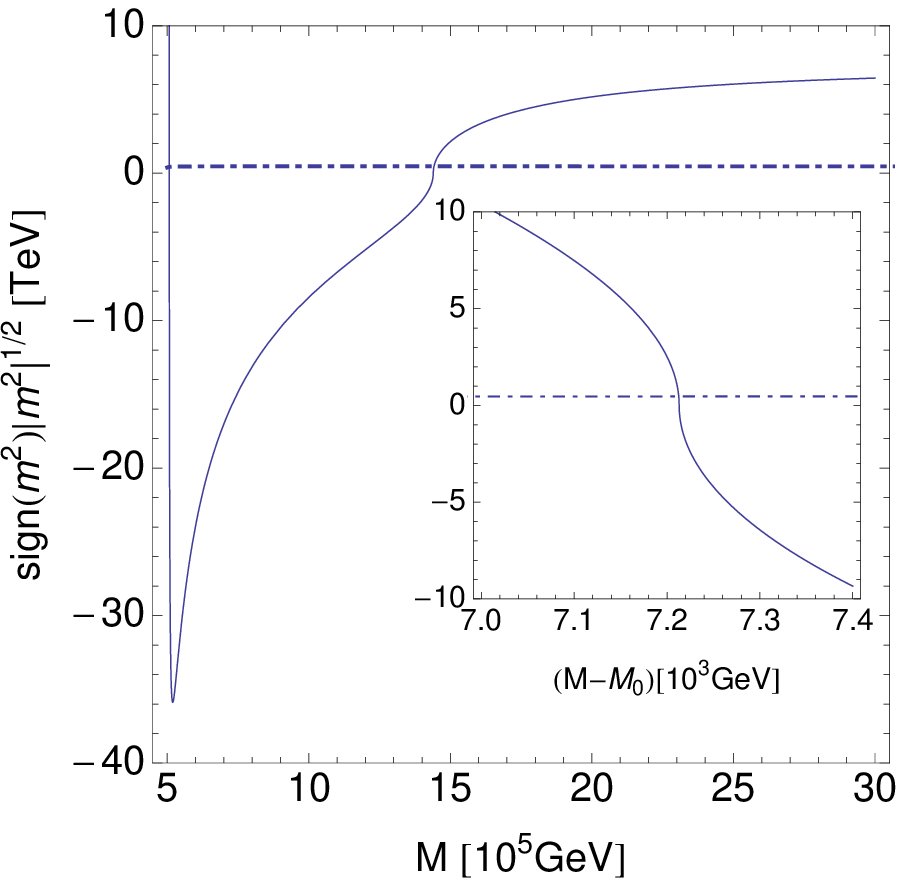}
\caption{Signed mass $\mathrm{sign}(m^2)\sqrt{|m^2|}$ of right-handed stau $\widetilde{\tau}_R$ as a function of the messenger scale $M$ for $h=0.9$ and fixed $Mx=5\times10^5\,\mathrm{GeV}$. The subplot displays the behaviour of signed mass for $M$ close to $M_0=5\times10^5\,\mathrm{GeV}$. The dot-dashed line represents mass of $\widetilde{\tau}_R$ in pure GMSB scenario (i.e. $h=0$).}\label{mseR3h09}
\end{center}
\end{figure}

\begin{figure}[h!]
\begin{center}
\includegraphics[scale=0.75]{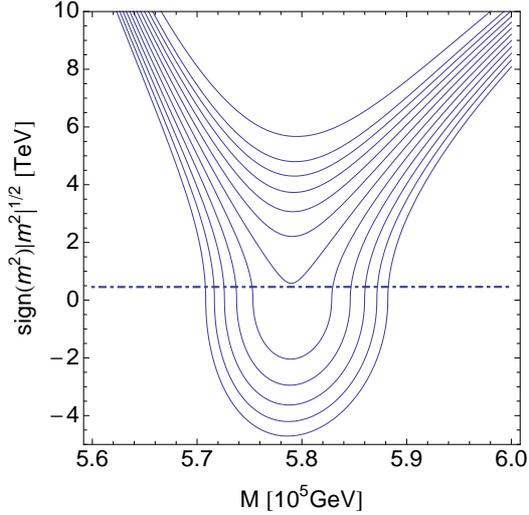}
\caption{Signed mass $\mathrm{sign}(m^2)\sqrt{|m^2|}$ of right-handed stau $\widetilde{\tau}_R$ as a function of the messenger scale $M$ for fixed $Mx=5\times10^5\,\mathrm{GeV}$ and $h=2.208$, $2.209$, \ldots, $2.219$, $2.220$ respectively, where the top-most line corresponds to $h=2.220$. The dot-dashed line represents mass of $\widetilde{\tau}_R$ in pure GMSB scenario (i.e. $h=0$).}\label{mseR3h2}
\end{center}
\end{figure}

To show the role which 1-loop function \eqref{fh1} and complete 2-loop function \eqref{fh2} play in the phenomenology of the EGMSB models we consider a specific example of \eqref{Wh} in which down-Higgs-type messenger superfield $Y_{H_d}$ couples to lepton superfields $L$, and $\overline{E}$:
\beq\label{Wexample}
W=h_{ij}Y_{H_d}L_i\overline{E}_j.
\eeq
We assume that $h_{ij}$ has similar hierarchy to the matrix of leptonic Yukawa couplings $(y_e)_{ij}$. One can check by direct computation that such superpotential gives extra contributions to $A_{\tau}$ and to the soft terms at 1- and 2-loop level only for sleptons:
\beqa
A_{\tau}&=&-\frac{h^2}{16\pi^2}3Mxf_{h,A}(x),
\eeqa
\beqa
m^2_{h,\widetilde{l}_3}&=&\frac{h^2}{16\pi^2}\left(-\frac{1}{6}\right)(Mx)^2x^2f_{h,1}(x)\nonumber\\
&&\left(\frac{h^2}{16\pi^2}\right)^24(Mx)^2f_{h,2}(x),\label{m2lL3}
\eeqa
\beqa
m^2_{h,\widetilde{\tau}_R}&=&\frac{h^2}{16\pi^2}\left(-\frac{1}{6}\right)2(Mx)^2x^2f_{h,1}(x)\nonumber\\
&&+\left(\frac{h^2}{16\pi^2}\right)^28(Mx)^2f_{h,2}(x)\label{m2eR3}.
\eeqa
These soft terms have been implemented in\footnote{Related \textsc{SARAH} file called EGMSB.m can be downloaded from \texttt{www.tjel.us.edu.pl/tools.html}. To generate \textsc{SPheno} code please use: \texttt{Start["MSSM"]} and then \texttt{MakeSPheno[InputFile$\rightarrow$"EGMSB.m"]}} \textsc{SARAH-4.5.8} \cite{Staub:2009bi,Staub:2010jh,Staub:2012pb,Staub:2013tta}, which we used to generate \textsc{SPheno-3.3.7} code \cite{Porod:2003um,Porod:2011nf}. The obtained data were further processed with the help of \textsc{SSP-1.2.1} \cite{Staub:2011dp}. For simplicity, in our analysis we fixed $\tan\beta=10$ and chose $Mx=5\times10^5\mathrm{GeV}$ such that the lightest Higgs mass $m_{h^0}$ is $123\,\mathrm{GeV}$, while masses of gluino $\widetilde{g}$ and 1st and 2nd generation squarks $\widetilde{q}_{1,2}$ are about $4-5\,\mathrm{TeV}$. For 
$h=0.1$, $0.9$ and $h\sim2.2$ we scanned mass spectrum over $M$ in the range $5\times10^5\,\mathrm{GeV}<M<5\times10^{9}\,\mathrm{GeV}$ what is equivalent to varying $x$ in the range $10^{-4}<x<1$.  As expected from \eqref{m2lL3} and \eqref{m2eR3} only 3rd generation slepton masses are sensitive to the $h$ coupling. Masses of other particles are similar to those in pure GMSB scenario:
\beqa
&&m_{H_i^0,A^0,H^{\pm}}\approx 1.7-1.8\,\mathrm{TeV},\\
&&m_{\widetilde{g}}\approx4.1\,\mathrm{TeV},\\
&&m_{\widetilde{\chi}^0_i}\approx0.8-1.7\,\mathrm{TeV},\quad m_{\widetilde{\chi}^{\pm}_i}\approx1.5-1.7\,\mathrm{TeV},\\
&&m_{\widetilde{u}_i}\approx 4.4-5.1\,\mathrm{TeV},\quad m_{\widetilde{d}_i}\approx4.8-5.1\,\mathrm{TeV},\\
&&m_{\widetilde{e}_{L},\widetilde{\mu}_L}\approx1.7\,\mathrm{TeV},\quad m_{\widetilde{\nu}_{e,\mu}}\approx1.7\,\mathrm{TeV},\\
&&m_{\widetilde{e}_{R},\widetilde{\mu}_R}\approx0.7\,\mathrm{TeV}.
\eeqa
Dependence of the signed mass of the right-handed stau $\widetilde{\tau}_R$ i.e. $\mathrm{sign}(m^2_{\widetilde{\tau}_R})\sqrt{|m^2_{\widetilde{\tau}_R}|}$ on the messenger mass $M$ for 
$h=0.1$ and $0.9$ is shown in Figs.~\ref{mseR3h01} and~\ref{mseR3h09} respectively. When one increases $h$ up to about $2.2$ then the dependence on $M$ changes such that $m^2_{\widetilde{\tau}_R}$ is positive for any value of $M$ -- see Fig.~\ref{mseR3h2}. 
As one can see the interplay between 1- and 2-loop contributions to soft masses discussed in Sec.~IV leads to large negative values of $m^2_{\widetilde{\tau}_R}$ when $M$ is close to $M_0=5\times10^5\,\mathrm{GeV}$. But when $M-M_0$ is of the order of $10^2-10^3\mathrm{GeV}$ then 2-loop contributions start to dominate and $m^2_{\widetilde{\tau}_R}$ is again positive. Such a behaviour is apparent especially for $h\sim2.2$ (see Fig.~\ref{mseR3h2}). One can see that in  this case $m^2_{\widetilde{\tau}_R}$ is quite sensitive to the precise value of $h$. The basic difference between discussed scenario and the standard GMSB mass pattern is that 3rd generation slepton masses can be as well of the order of $500\,\mathrm{GeV}$ (as in the pure GMSB scheme) as much larger i.e. of the order $10-20\,\mathrm{TeV}$. In addition, contrary to the pure GMSB case, for the fixed value of $Mx$ slepton masses heavily depend on the the value of $M$. 
Here they can even be tachyonic what could lead to instability of the scalar potential.  Finally, let us note that 
for $M>10^8\,\mathrm{GeV}$, what is the most common choice of messenger scale, \eqref{m2lL3} and \eqref{m2eR3} dominate over GMSB contributions making masses of 3rd generation of sfermions of the order of $10\,\mathrm{TeV}$. Such a hierarchy in the slepton spectrum is a simple consequence of choosing \eqref{Wexample} similar to the MSSM leptonic Yukawa couplings.  As one can see complete 2-loop contributions to soft masses considered here have a large impact on the 3rd generation of sleptons.

\section{Summary and Outlook}

In this letter, we have presented full 2-loop calculation in the EGMSB model with one coupling between  messenger and matter. We have shown that full 2-loop analytical contributions to soft masses are important in the regime $x=F/M^2\sim1$, which is realized e.g. in the Low-Scale Gauge Mediation models with messenger mass $M\sim10^5-10^6\mathrm{GeV}$.  
It turned out that for $x\sim1$ the model exhibits novel behaviour with respect to the standard GMSB case.
It has been shown on a specific example that considered here complete 2-loop effects can significantly affect slepton spectrum, 
what opens up 
new scenarios beyond the standard GMSB scheme.
%
%
%
%

Let us shortly comment on the remaining contributions from higher loops. Taking into account suppression of 1-loop contribution with respect to 2-loop contribution, the soft mass $m^2_h$ can be written in the following form:
\beqa\label{3loop}
\frac{m^2_h}{(Mx)^2}&=&\frac{h^2}{16\pi^2}k_1x^2f_{1}(x)+\left(\frac{h^2}{16\pi^2}\right)^2k_2f_{2}(x)\nonumber\\
&&+\left(\frac{h^2}{16\pi^2}\right)^3k_3x^{n_3}f_{3}(x)+\ldots,
\eeqa
where $k_1=-1/6$, $k_2=3$, $k_3$ is unknown numerical factor of 3-loop contribution, while $f_{i}(x)$ are 
functions normalized such that $f_{i}(0)=1$. Finally 
$n_3\geq0$ is unknown power of $x$ which appears in the 3-loop contribution. 
We are not aware of any 
strong argument that $n_3=0$ and
$f_3(x)$ is of the order 1 for $x\lesssim0.9$,   
but from the fact that in \eqref{3loop} the suppression by $x^2$ occurs  only at 1-loop and not at 2-loop level, one might conjecture that $n_3=0$. We also expect that the function $f_3(x)$ behaves similarly to $f_2(x)$ i.e. it is of the order of $1-10$ for $x\lesssim0.9$, what in turn could suggests that the ratio of 3- to 2-loop contribution is of the order of
$(k_3/k_2)h^2/16\pi^2\lesssim0.1$. 
Certainly, from the theoretical point of view it would be interesting to obtain even approximate 3-loop result and verify those assumptions.   


\section*{Acknowledgements}

TJ would like to thank Micha{\l} Czakon for some useful technical comments and discussions. This work was supported by the Polish National Science Centre (NCN) under the Grant Agreement No. DEC-2013/11/B/ST2/04023 and under postdoctoral grant No. DEC-2012/04/S/ST2/00003.

\section*{Appendix}

Here we collect contributions to self-energy \eqref{Pi2} from topologies $(W,S,\ldots)$ defined on Fig.~2 in \cite{Martin:2003it}. We use the following notation: $z_{\pm}=M^2(1\pm x)$, while $c=m^2_{\mathrm{IR}}$ regulates IR divergencies  in 2-loop integrals at $s=0$. The sum of \eqref{PiWSSSS}-\eqref{PiVFFFFS} is finite in the limit $c\to0$. The components of \eqref{Pi2} are 
expressed in terms of functions $(W_{SSSS}, X_{SSS}, Y_{SSSS}, \ldots)$ defined in the Sec. IV.A and IV.B of \cite{Martin:2003it}. 

\beqa\label{PiWSSSS}
&&\frac{4}{h^4z}\Pi_{W_{SSSS}}=-2 W_{SSSS}(z_+,z_-,c,c)\nonumber\\
&&+2 W_{SSSS}(c,c,c,z_-)+W_{SSSS}(z_-,z_-,c,c)\nonumber\\
&&+2W_{SSSS}(c,c,c,z_+)+W_{SSSS}(z_+,z_+,c,c),
\eeqa

\beqa
&&\frac{2}{h^4}\Pi_{X_{SSS}}=X_{SSS}(z_-,z_+,c)\nonumber\\
&&+X_{SSS}(z_+,z_-,c)+X_{SSS}(c,c,z_-)\nonumber\\
&&+X_{SSS}(z_-,z_-,c)+X_{SSS}(c,c,z_+)\nonumber\\
&&+X_{SSS}(z_+,z_+,c)+2 X_{SSS}(c,c,c),
\eeqa

\beqa
&&\frac{4}{h^4z}\Pi_{Y_{SSSS}}=-2 Y_{SSSS}(c,z_-,z_+,c)\nonumber\\
&&-2Y_{SSSS}(c,z_+,z_-,c)+Y_{SSSS}(z_-,c,c,z_+)\nonumber\\
&&+Y_{SSSS}(z_+,c,c,z_-)\nonumber\\
&&+2 Y_{SSSS}(c,z_-,z_-,c)+2 Y_{SSSS}(z_-,c,c,c)\nonumber\\
&&+Y_{SSSS}(z_-,c,c,z_-)+2Y_{SSSS}(c,z_+,z_+,c)\nonumber\\
&&+2 Y_{SSSS}(z_+,c,c,c)+Y_{SSSS}(z_+,c,c,z_+),
\eeqa

\beqa
&&\frac{16}{h^4z}\Pi_{Z_{SSSS}}=-Z_{SSSS}(c,z_-,c,z_+)\nonumber\\
&&-Z_{SSSS}(c,z_-,z_+,c)-Z_{SSSS}(c,z_+,c,z_-)\nonumber\\
&&-Z_{SSSS}(c,z_+,z_-,c)-Z_{SSSS}(z_-,c,c,z_+)\nonumber\\
&&-Z_{SSSS}(z_-,c,z_+,c)-Z_{SSSS}(z_+,c,c,z_-)\nonumber\\
&&-Z_{SSSS}(z_+,c,z_-,c)+Z_{SSSS}(c,z_-,c,z_-)\nonumber\\
&&+Z_{SSSS}(c,z_-,z_-,c)+Z_{SSSS}(z_-,c,c,z_-)\nonumber\\
&&+Z_{SSSS}(z_-,c,z_-,c)+Z_{SSSS}(c,z_+,c,z_+)\nonumber\\
&&+Z_{SSSS}(c,z_+,z_+,c)+Z_{SSSS}(z_+,c,c,z_+)\nonumber\\
&&+Z_{SSSS}(z_+,c,z_+,c),
\eeqa

\beqa
&&\frac{4}{h^4}\Pi_{S_{SSS}}=S_{SSS}(c,z_-,z_+)\nonumber\\
&&+S_{SSS}(c,z_+,z_-)+S_{SSS}(c,z_-,z_-)\nonumber\\
&&+S_{SSS}(c,z_+,z_+)+4 S_{SSS}(c,c,c),
\eeqa

\beqa
&&\frac{4}{h^4z}\Pi_{U_{SSSS}}=-U_{SSSS}(z_-,c,c,z_+)\nonumber\\
&&-U_{SSSS}(z_-,c,z_+,c)-U_{SSSS}(z_+,c,c,z_-)\nonumber\\
&&-U_{SSSS}(z_+,c,z_-,c)+4U_{SSSS}(c,z_-,c,c)\nonumber\\
&&+U_{SSSS}(z_-,c,c,z_-)+U_{SSSS}(z_-,c,z_-,c)\nonumber\\
&&+4U_{SSSS}(c,z_+,c,c)+U_{SSSS}(z_+,c,c,z_+)\nonumber\\
&&+U_{SSSS}(z_+,c,z_+,c),
\eeqa

\beqa
&&\frac{4}{h^4z^2}\Pi_{V_{SSSSS}}=V_{SSSSS}(c,z_-,z_+,c,c)\nonumber\\
&&+V_{SSSSS}(c,z_+,z_-,c,c)+V_{SSSSS}(z_-,c,c,c,z_+)\nonumber\\
&&+V_{SSSSS}(z_+,c,c,c,z_-)+V_{SSSSS}(c,z_-,z_-,c,c)\nonumber\\
&&+V_{SSSSS}(z_-,c,c,c,z_-)+V_{SSSSS}(c,z_+,z_+,c,c)\nonumber\\
&&+V_{SSSSS}(z_+,c,c,c,z_+),
\eeqa

\beqa
&&\frac{4}{h^4}\Pi_{W_{SFF}}=4 W_{SFF}(c,c,0,z)\nonumber\\
&&+W_{SFF}(z_-,z_+,0,0)+W_{SFF}(z_+,z_-,0,0)\nonumber\\
&&+W_{SFF}(z_-,z_-,0,0)+W_{SFF}(z_+,z_+,0,0),
\eeqa

\beqa
&&\frac{4}{h^4z}\Pi_{V_{SSSFF}}=2 V_{SSSFF}(z_-,c,c,0,z)\nonumber\\
&&+2 V_{SSSFF}(z_+,c,c,0,z)-V_{SSSFF}(c,z_-,z_+,0,0)\nonumber\\
&&-V_{SSSFF}(c,z_+,z_-,0,0)+V_{SSSFF}(c,z_-,z_-,0,0)\nonumber\\
&&+V_{SSSFF}(c,z_+,z_+,0,0),
\eeqa

\beqa\label{PiVFFFFS}
&&\frac{2}{h^4}\Pi_{V_{FFFFS}}=4 V_{FFFFS}(0,z,z,0,c)\nonumber\\
&&+2V_{FFFFS}(z,0,0,z,c)+V_{FFFFS}(z,0,0,0,z_-)\nonumber\\
&&+V_{FFFFS}(z,0,0,0,z_+).
\eeqa

The complete analytical form of the function $f_{h,2}(x)$ which enters the 2-loop contribution to soft mass \eqref{m2h2} is:

\beqa\label{fh2}
&&f_{h,2}(x)=\nonumber\\
&&\frac{1}{72x^2(1-x^2)^2}\bigg[12 (x+1)^2 \left(x^2-2 x+2\right) x^2 \text{Li}_2(1-x)\nonumber\\
&&-12 (x-1)^2 \left(x^2+2 x+2\right) x^2\text{Li}_2\left(\frac{1}{x+1}\right)\nonumber\\
&&-24 \left(x^2+1\right) x^3 \text{Li}_2\left(\frac{1-x}{1+x}\right)+48 x^5 \ln(2 x) \mathrm{artanh}(x)\nonumber\\
&&-12 x^4 \ln x \ln (x+1)-24 (x^2-1) (x-1) x^3 \zeta(2)\nonumber\\
&&+48 x^3 \ln (2) \mathrm{artanh}(x)+24(x^2-1) x^2\nonumber\\
&&-6 (x^2-1) \left(19 x^2-20\right) \ln\left(1-x^2\right)\nonumber\\
&&-12x(x^2-1) \left(7 x^2-10\right) \mathrm{artanh}(x)\nonumber\\
&&+12 \left(x^4+2\right) x^2 \ln x \ln\left(1-x^2\right)\nonumber\\
&&-3 (x+1)^2 \left(4 x^4-5 x^3-10 x^2+31x-16\right) \ln ^2(x+1)\nonumber\\
&&+12\ln (1-x)\left[x^2 \left(x^3-x^2+x+3\right) \ln (x+1)-x^4 \ln x\right]\nonumber\\
&&-3(x-1) \left(2 x^5+3 x^4-11 x^3-17 x^2+15 x+16\right) \ln ^2(1-x)\bigg]\nonumber\\
&&-\frac{1}{6x^2(1-x^2)}(2-x^2)\left[x\,\mathrm{arctanh}(x)+\ln(1-x^2)\right].
\eeqa
 This function can be found altogether with some additional material in \texttt{www.tjel.us.edu.pl/tools.html}.


\end{document}